\documentclass[%
 reprint,
showkeys,
 amsmath,amssymb,
 aps,
]{revtex4-1}

\usepackage{graphicx}
\usepackage{dcolumn}
\usepackage{bm}

\renewcommand{\vec}[1]{\mbox{\boldmath $#1$}}

\newcommand{\upd}{\mathrm{d}}

\begin{document}

\preprint{APS/123-QED}

\title{
Quantifying hierarchical mixture quality in polymer composite materials:
structure and inhomogeneity in multiple scales
}%

\author{Yasuya Nakayama}
 \email{nakayama@chem-eng.kyushu-u.ac.jp}
\author{Toshihisa Kajiwara}
\affiliation{%
Department of Chemical Engineering,
Kyushu University,
Nishi-ku,
Fukuoka 819-0395,
Japan
}%


\date{Dec 21, 2012}

\begin{abstract}
Mixture quality plays a crucial role in the physical properties of
multi-component immiscible polymer mixtures including
nanocomposites and polymer blends.
Such complex mixtures are often characterized by hierarchical internal
 structures, which have not been accounted for by conventional mixture quantifications.
We propose a way to characterize the mixture quality of complex mixtures
 with hierarchical structures.
Starting from a concentration field, which can be typically obtained
 from TEM/SEM images, the distribution of the coarse-grained concentration
 is analyzed to obtain the scale-dependent inhomogeneity of a mixture.
The hierarchical nature of a mixture is characterized by multiple
 characteristic scales of the scale-dependent inhomogeneity.
We demonstrate how the proposed method works to characterize 
sizes and distributions in different dispersions.
This method is generally applicable to various complex mixtures.
%
\end{abstract}

\maketitle


\section{Introduction}
Multi-component polymer materials including nanocomposites,
fiber-reinforced plastics, and polymer blends typically have 
inhomogeneous internal structures on a mesoscopic level.
The properties of such mixture materials are strongly dependent on
the distribution of their internal structures, namely, the mixture quality
\cite{Robeson2007Polymer,Schaefer2007How,Vaia2007Polymer,Paul2008Polymer,Gibson2010Review}.
Therefore, there is a common need to develop a quantification measure to
describe the mixture quality of multi-component polymer materials.
Such a technique would also be required to evaluate mixing processes in
various mixing devices in the field of chemical engineering
\cite{tadmor06:_princ_of_polym_proces,cullen09:_food_mixin,muthukumarappan09:_mixin_fundam,Ottino1983Laminar,Bigio1990Measures,Stone2005Imaging,Funakoshi2008Chaotic,Bothe2006Fluid,Bothe2008Computation}.
Though various methods for the analysis of mixture quality have been
proposed, these approaches are sometimes system-specific and fail to
directly assess hierarchical internal structures.
To compare different mixture systems (different filler loading,
different mixing conditions, etc.), a simple quantification method based
on non-system-specific criteria is required.

In general, internal structures in a complex mixture have their own
characteristic scales. 
Therefore, the simultaneous quantification of size and distribution is
necessary.
However, deviation from uniformity, which is a commonly used
method, often misses characteristic scales of internal inhomogeneity
because uniform distribution itself is intrinsically defined without any scale.
From this observation, the identification of characteristic scales is a
fundamental problem when discussing the mixture quality of complex
mixtures.

Many prior works regarding the quantification of mixture quality
primarily focused on the non-uniformity of distribution, regardless of
the sizes or structures of the dispersed phase.
Typically for a dispersion of spherical particles, 
the center points of the particles are first identified, then the distribution
of the center points is analyzed
\cite{Zhu2010Statistical,Li2012Automatic,Sul2011Quantitative}.
One direct approach to measure the distribution inhomogeneity is to
evaluate the deviation from the uniform distribution based on
fluctuation of the point density
\cite{yang94:_flow_field_analy_of_banbur_mixer,yang09:_flow_field_analy_of_banbur_mixer,connelly07:_examin_of_mixin_abilit_of,alemaskin05:_color_mixin_in_meter_zone,camesasca09:_danck_revis_use_of_entrop,Zhu2010Statistical,Li2012Automatic}.
Another approach is to evaluate a certain cost function of the
inter-point distance, which is the minimum when the points are uniformly
distributed~\cite{Sul2011Quantitative}.
These techniques are useful for systems where the ideal distribution is
homogeneous but not for systems with internal structures.
The length scales of internal structures cannot assessed by these approaches.

Instead of measuring deviation from uniformity, a degree of
clustering was used to characterize the inhomogeneity of the distribution.
The spatial correlation function of density of the dispersed phase is
the simplest way to define correlation length, which is an average size
of the cluster
\cite{danckwerts52:_defin_and_measur_of_some,muthukumarappan09:_mixin_fundam}.
Alternative methods to define size were developed based on either a
concentration gradient~\cite{Bothe2006Fluid}, or the inter-point
distance~\cite{Stone2005Imaging}.
The size of the matrix phase not containing the dispersed phase can also
be used to characterize the clustering tendency, which is closely
related to the mechanical reinforcement of
nanocomposites~\cite{Khare2010Quantitative}.
Another method is based on the mathematical morphology, 
which considers the volume fraction of isotropic virtual dilation of the dispersed phase.
The dependence of the volume fraction on dilation differs depending on
the degree of clustering~\cite{Pegel2009Spatial}.
These approaches focus on one representative length scale but not on
the multiple characteristic scales associated with hierarchical internal
structures.

Internal structures within a complex mixture are often hierarchical in
nature.
Thus, different scales associated with hierarchical structures should be
involved in characterizing the mixture quality of a complex mixture.
The only other known study that has assessed inhomogeneity at different
scales involves the Mix-Norm method~\cite{Mathew2005Multiscale}.
However, the Mix-Norm method has focuses on the defining non-uniformity of a system
in which the ideal mixture state is uniform for all scales and thus does not
apply to the identification of hierarchical characteristic scales.
In this paper, we propose a scale-dependent measure for inhomogeneity of
complex mixtures that is capable of identifying multiple characteristic
scales of inhomogeneity.
The degrees of inhomogeneity are defined at different scales through
the coarse-graining of the concentration field of a dispersed phase.
Multiple characteristic scales are identified with the scale dependence
of inhomogeneity.
We demonstrate how the scale-dependent measure works by applying the
technique to synthetic dispersions with different distributions.
This technique can aid in comparing different complex mixtures and can
provide a fundamental understanding of the mixture quality of systems
with different internal structures.

\section{Method of scale-dependent moment for mixture quality}
We start with a concentration field, \(c(\vec{x})\), of a component in a
mixture.
The inhomogeneity of the mixture can be characterized based on the
fluctuation in the concentration field, which contains information on a
microscopic level. The resolution of \(c(\vec{x})\) is determined by the
resolution of the data obtained.

Because the definition of inhomogeneity should not depend on the total
content of the component, we consider the normalized concentration
field, namely a probability density, as
\begin{align}
 \mu(\vec{x})&=\frac{c(\vec{x})}{\int_{\Omega}\upd\vec{y}c(\vec{y})},
\end{align}
where \(\Omega\) denotes the whole domain of the mixture. The induced
probability density \(\mu(\vec{x})\) satisfies \(
\int_{\Omega}\upd\vec{x}\mu(\vec{x})=1 \) irrespective of the total
content of the dispersed substance
\(\int_{\Omega}\upd\vec{x}c(\vec{x})\).

To quantify the fluctuation of a specific scale, the coarse-grained field
is defined as
\begin{align}
 \mu_{r}(\vec{x})&=\int_{B_{r}(\vec{x})}
 \!\!\!\!\!\!
 \!\!\!
 \upd\vec{y}\mu(\vec{x}+\vec{y}),
\label{eq:coarse-grained_measure}
\end{align}
where \(B_{r}(\vec{x})\) denotes the domain of a linear size \(r\)
around a location \(\vec{x}\).
Coarse-graining is a familiar concept in statistical physics and has
been especially useful in characterizing fractal structure in fluid
turbulence
\cite{kolmogorov62:_refin_of_previous_hypot_concer,oboukhov62:_some_specif_featur_of_atmos_tubul,mandelbrot82:_fract_geomet_of_natur,fujisaka03:_inter_and_expon_field_dynam}.
Coarse-graining has also been applied to characterize time-series with
long-range correlation\cite{fujisaka87:_inter_caused_by_chaot_modul}.
In these applications, fractal structures were characterized by power-law scaling,
and the exponents of this analysis were a primary concern.
Here, we are interested in applying the coarse-graining concept to
quantification of mixture quality.

The fluctuation at a resolution scale \(r\) is characterized by the
statistical moment,
\begin{align}
 \langle \mu_{r}^{q}\rangle &\equiv \int\upd\vec{x}\mu_{r}^{q}(\vec{x}),
 ~q\in(-\infty, \infty),
\label{eq:q_moment}
\end{align}
where the parameter \(q\) represents the magnitude of fluctuation: a large 
positive \(q\) indicates a large fluctuation and a negative \(q\) indicates
a small fluctuation.

If a mixture is homogeneous at a scale \(r\),
\begin{align}
 \langle \mu_{r}^{q}\rangle^{p} &=
 \langle \mu_{r}^{p}\rangle^{q}~~\text{for}~~q\neq p,
\label{eq:mu_qp}
\end{align}
holds.
For instance, in the limit of macroscopic scale, \(r\to L\), where \(L\)
denotes the system size, and the coarse-grained concentration becomes
\(\mu_{L}=\int_{B_{L}}\upd\vec{x}\mu(\vec{x})\approx
\int_{\Omega}\upd\vec{x}\mu(\vec{x})\) by definition; therefore, the
normalized moment is always close to unity at the system size,
\(\langle\mu_{L}^{q}\rangle/\langle \mu_L\rangle^{q}\approx 1\).
This property simply represents the macroscopic homogeneity in the general situation.
Furthermore, if there is no coherent structure of scale \(r\), the
relationship
\begin{align}
 \langle \mu_{r}^{q}\rangle
 &\sim A r^{Dq},\label{eq:scaling1}
\end{align}
holds with a coefficient \(A\), where \(D\) is the spatial dimension. 
Based on the properties (\ref{eq:mu_qp}) and (\ref{eq:scaling1}), to
characterize the inhomogeneity of a mixture, we define the {\itshape
inhomogeneity function at scale \(r\)} by a normalized moment,
\begin{align}
 F_{q}(r) &= 
\left[
\frac{
 \langle \mu_{r}^{q}\rangle
 }{
 \langle \mu_{r}\rangle^{q}
 }
\right]^{1/(q-1)}.
\end{align}

To understand the physical implication of \(F_{q}(r)\),
we consider a mixture with a correlation length \(l\).
In this case, the inhomogeneity function behaves as
\begin{align}
F_{q}(r) &
 \begin{cases}
 \to \text{cst.}(>1)& r\ll l,
  \\
  > 1
  & r \sim l,
  \\
  \to  1
  & r\gg l,
 \end{cases}
\label{eq:fqr_scale_l}
\end{align}
The fluctuation in a smaller scale than \(r\) is accounted for in
\(F_{r}(q)\).
Three ranges of characteristic scales are observed.
In the large-scale
regime of \(r\gg l\),
the mixture appears to be homogeneous.
This leads that Eq.(\ref{eq:mu_qp}) holds, and then \(F_{q}(r)=1\) is obtained
irrespective of \(q\).
At the small-scale regime of \(r\ll l\), the mixture has a certain level
of inhomogeneity.
In this scale, Eq.(\ref{eq:scaling1}) holds and \(F_{q}(r)\sim
\left(A/A^{q}\right)^{1/(q-1)}=A^{-1}=\text{cst.}\) is obtained
asymptotically irrespective of \(q\).
Thus, \(F_{q}(r)\) approaches a certain level depending
on non-uniformity at the smallest scale.
At the correlation scale of \(r\sim l\), the rapid variation of
\(F_{q}(r)\) characterizes the correlation scale in the mixture.
From the inhomogeneity function, we can identify both the macroscopic
homogeneity and microscopic inhomogeneity of a mixture.
To characterize macroscopic homogeneity, we define the scale of
homogeneity, \(l_H\), which is a lower bound, as \(F_{q}(r)\approx 1\)
holds for \(l_{H}<r\).
In general, \(F_{q}(r)\) is a monotonously decreasing function of the
scale \(r\), which corresponds to the decreasing inhomogeneity as the
observation scale enlarges.
To identify a characteristic scale in a mixture clearly, it is
convenient to see the first derivative of \(F_{q}(r)\). We define the
{\itshape scale susceptibility} of the concentration fluctuation as
\begin{align}
 G_{q}(r)&\equiv -\frac{\upd F_{q}(r)}{\upd \ln r},
\end{align}
which takes a finite positive value when \(F_{q}(r)\) decreases sharply
based on the characteristic scale of
the fluctuation.
Because we are foucusing on hierarchical structures with a wide range of scales, 
the
scale susceptibility 
is defined as the derivative with respect to \(\ln r\).

\begin{figure}[htbp]
\begin{tabular}{|p{3.5em}||c|c|c|c|}
 \hline
 & \multicolumn{4}{c}{scale of segregation} \\\cline{2-5}
 \begin{minipage}[cbt]{\hsize}
  intensity ratio
  \vspace*{1ex}
 \end{minipage}
 &
 \(l=\)84
 & 
 \(l=\)28
 & 
 \(l=\)7
 & 
 \(l=\)2
 \\
\hline
\cline{2-5}
\cline{2-5}
0:255
 &
 \begin{minipage}{.18\hsize}
  \includegraphics[width=1.\hsize]{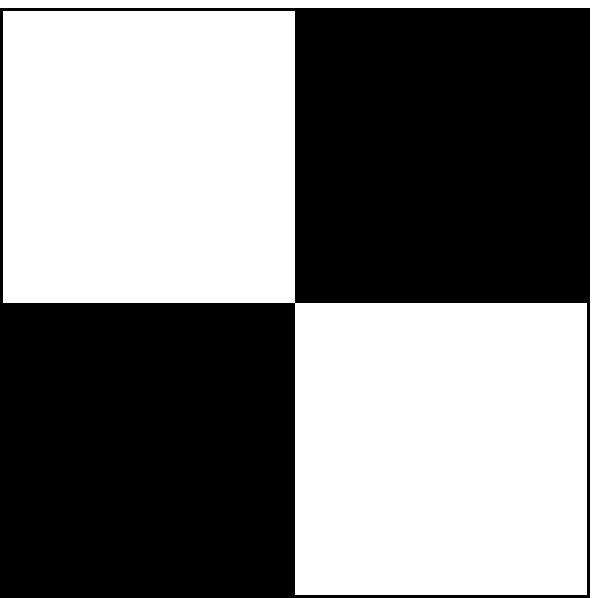}
 \end{minipage}
 &
 \begin{minipage}{.18\hsize}
  \includegraphics[width=1.\hsize]{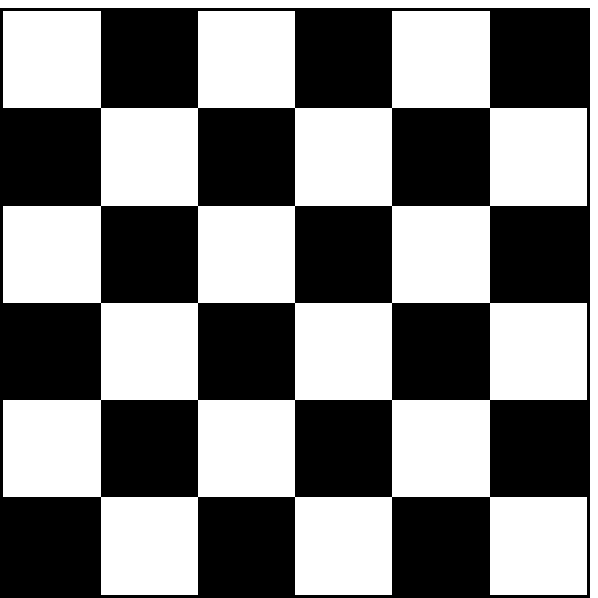}
 \end{minipage}
 &
 \begin{minipage}{.18\hsize}
  \includegraphics[width=1.\hsize]{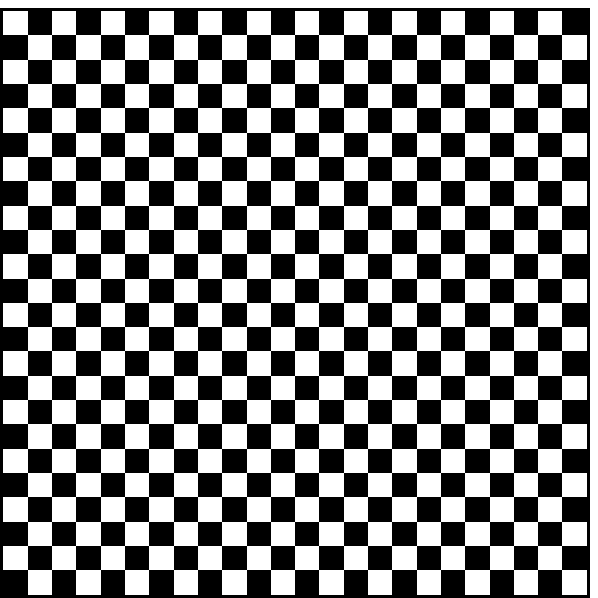}
 \end{minipage}
 &
 \begin{minipage}{.18\hsize}
  \includegraphics[width=1.\hsize]{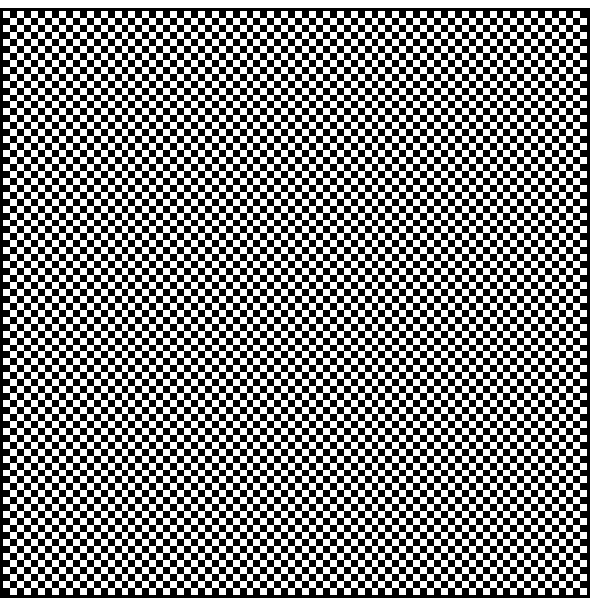}
  \end{minipage}
  \\
\hline
31:224 &
  \begin{minipage}{.18\hsize}
  \includegraphics[width=1.\hsize]{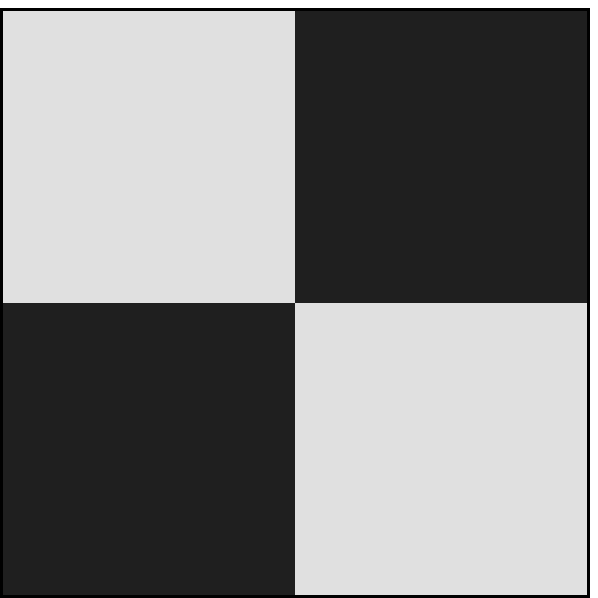}
  \end{minipage}
  &
  \begin{minipage}{.18\hsize}
  \includegraphics[width=1.\hsize]{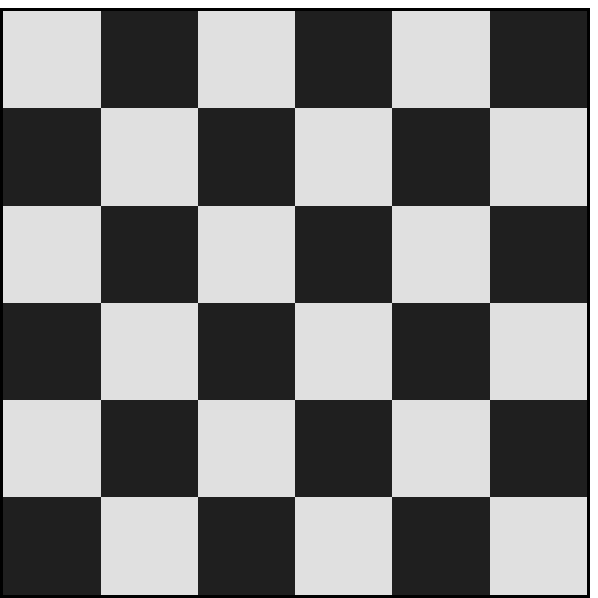}
  \end{minipage}
  &
  \begin{minipage}{.18\hsize}
  \includegraphics[width=1.\hsize]{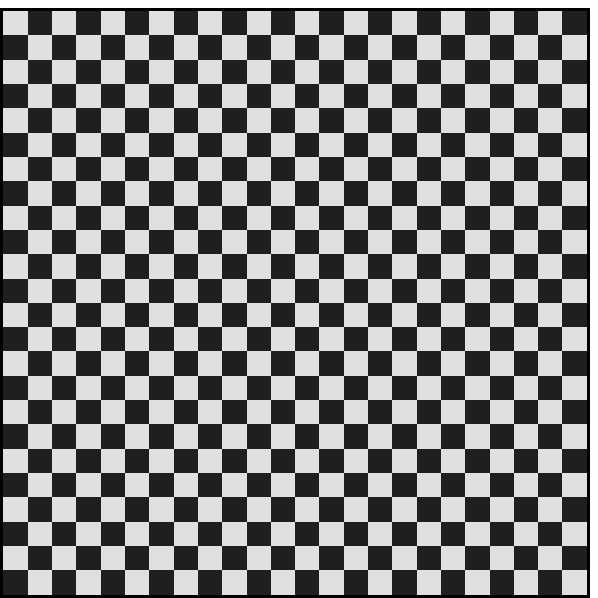}
  \end{minipage}
  &
  \begin{minipage}{.18\hsize}
  \includegraphics[width=1.\hsize]{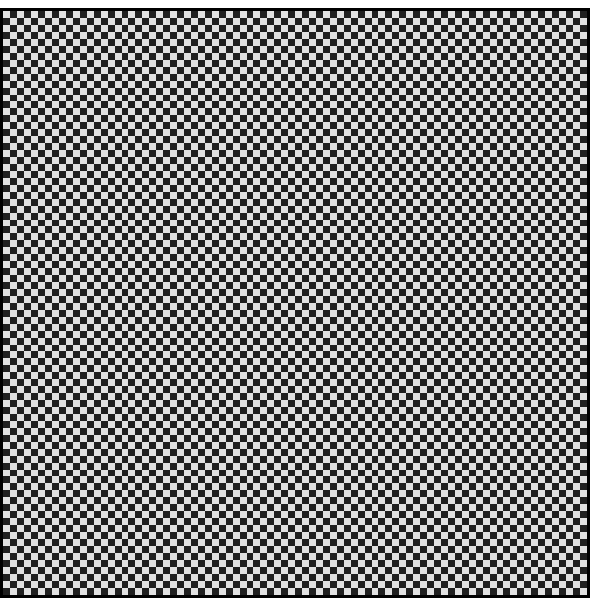}
  \end{minipage}
	     \\
\hline
63:192 &
 \begin{minipage}{.18\hsize}
  \includegraphics[width=1.\hsize]{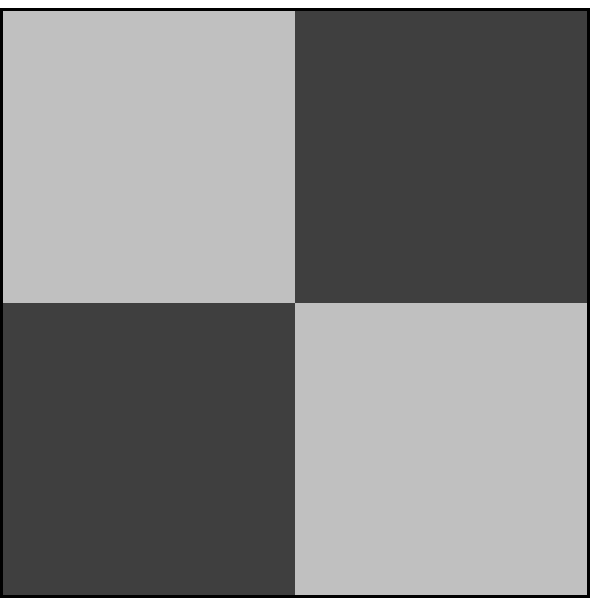}
 \end{minipage}
 &
   \begin{minipage}{.18\hsize}
  \includegraphics[width=1.\hsize]{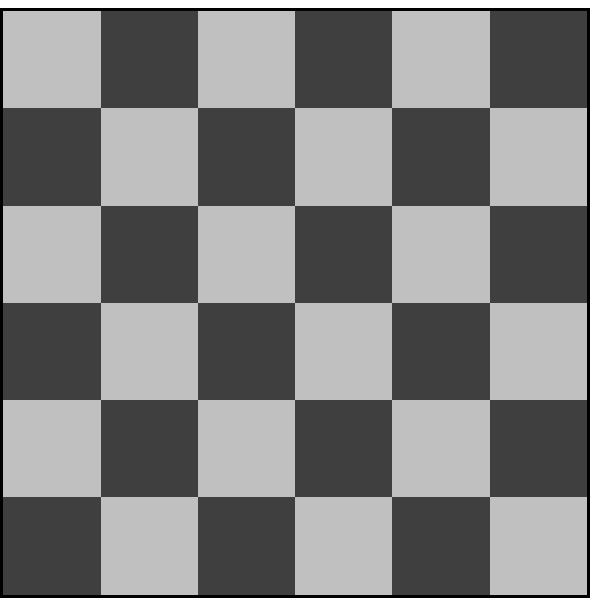}
   \end{minipage}
 &
 \begin{minipage}{.18\hsize}
  \includegraphics[width=1.\hsize]{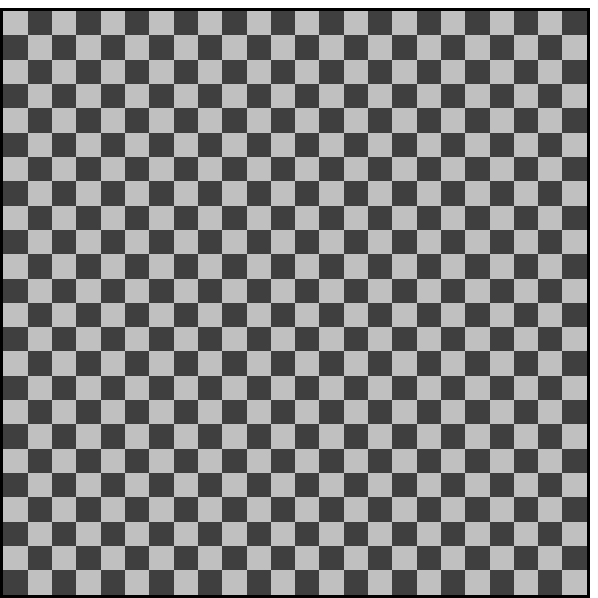}
 \end{minipage}
 &
 \begin{minipage}{.18\hsize}
  \includegraphics[width=1.\hsize]{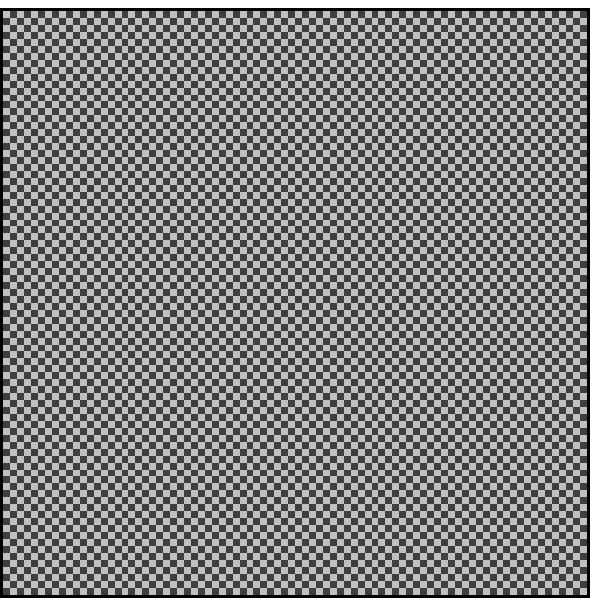}
 \end{minipage}
 \\
 \hline
 95:160
 &
 \begin{minipage}{.18\hsize}
  \includegraphics[width=1.\hsize]{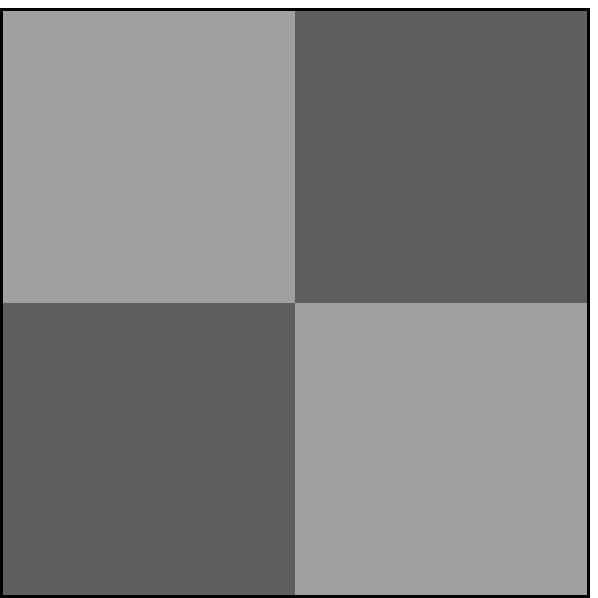}
 \end{minipage}
     &
 \begin{minipage}{.18\hsize}
  \includegraphics[width=1.\hsize]{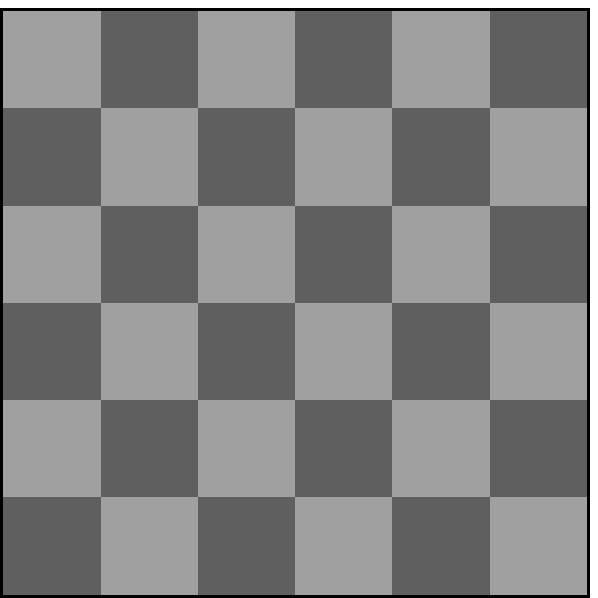}
 \end{minipage}
 &
 \begin{minipage}{.18\hsize}
  \includegraphics[width=1.\hsize]{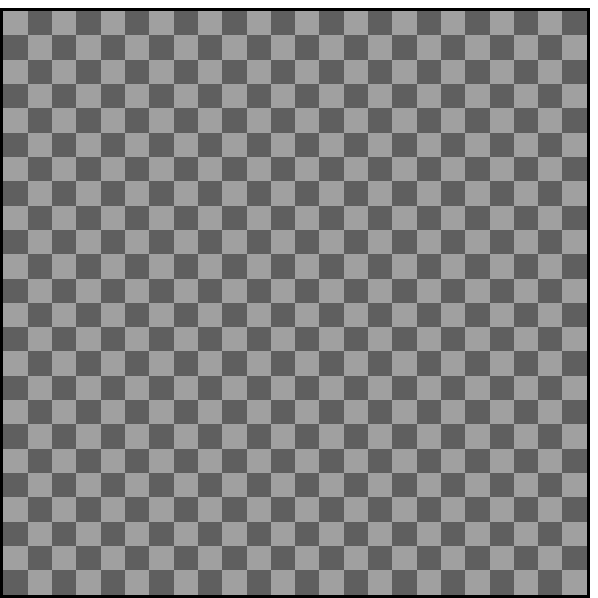}
 \end{minipage}
 &
 \begin{minipage}{.18\hsize}
  \includegraphics[width=1.\hsize]{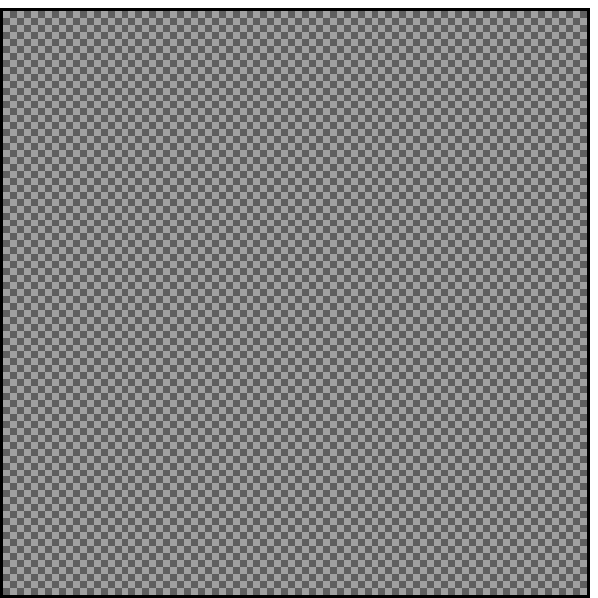}
 \end{minipage}
 \\
 \hline
 127:128 
 &
     \begin{minipage}{.18\hsize}
  \includegraphics[width=1.\hsize]{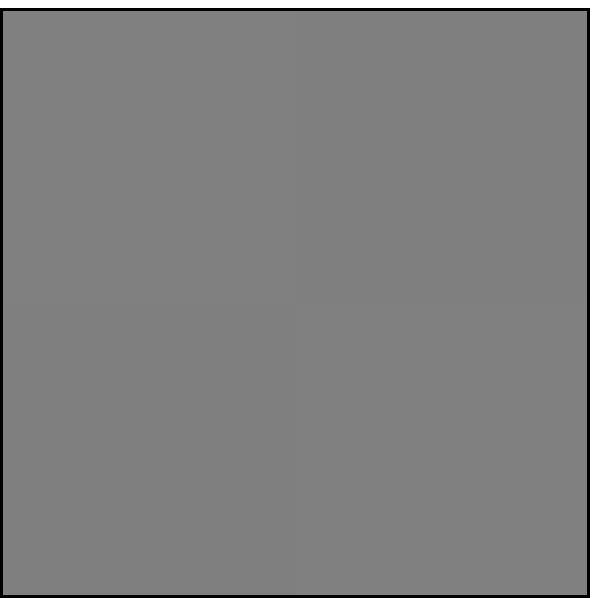}
     \end{minipage}
     &
 \begin{minipage}{.18\hsize}
  \includegraphics[width=1.\hsize]{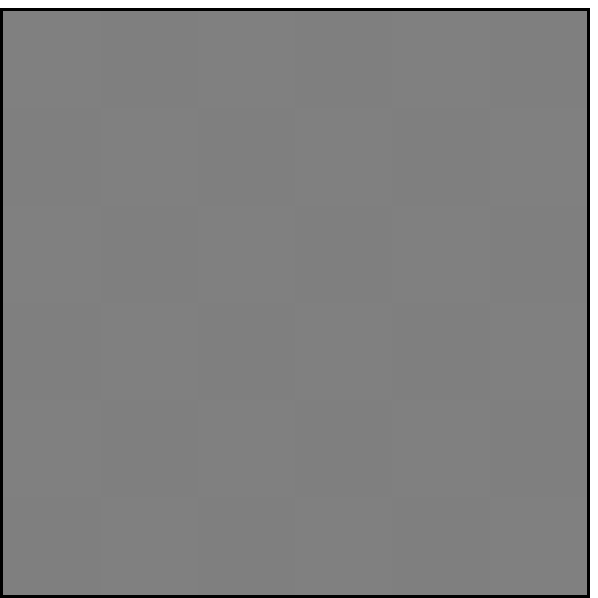}
 \end{minipage}
 &
\begin{minipage}{.18\hsize}
  \includegraphics[width=1.\hsize]{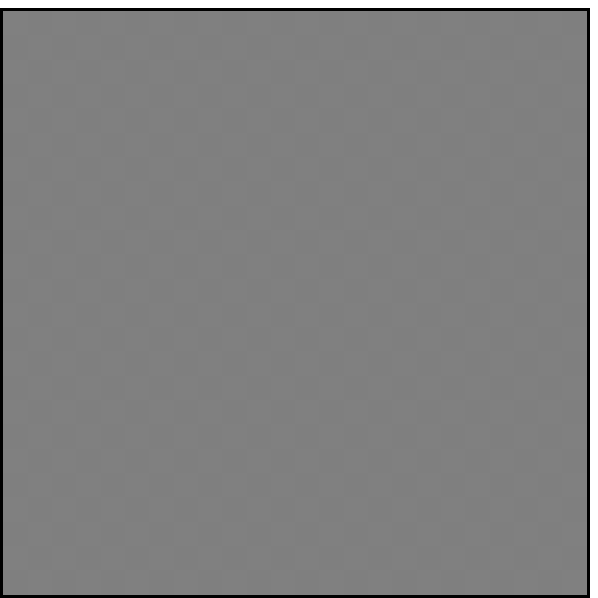}
\end{minipage}
 &
 \begin{minipage}{.18\hsize}
  \includegraphics[width=1.\hsize]{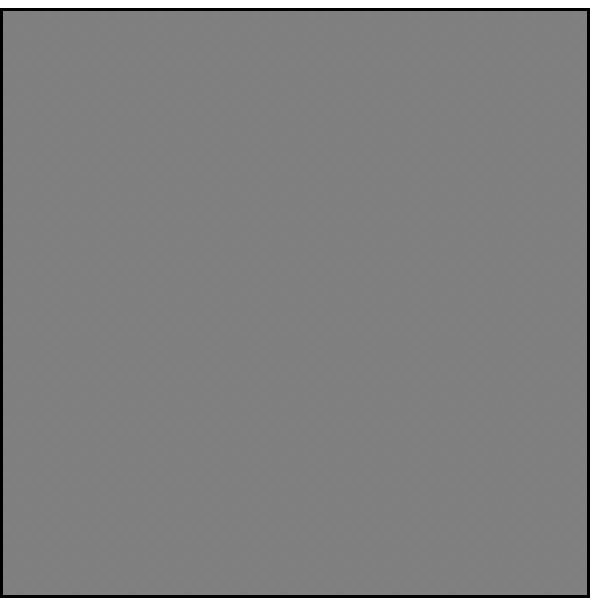}
 \end{minipage}
\\
\hline
\end{tabular}
\caption{Model systems of partially miscible blends of
 168\(\times\)168 pixels:
checkerboard patterns with varying shades of gray and varying
characteristic scales of 84\(\times\)84, 28\(\times\)28, 12\(\times\)12,
7\(\times\)7, and 2\(\times\)2 pixels.
The intensities of black and white regions are the same in each row and
are indicated in the first column.  }
\label{fig:dichotomous_tonedown_pattern}
\end{figure}

For the cases described in Eq.(\ref{eq:fqr_scale_l}), the scale susceptibility
becomes
\begin{align}
 G_{q}(r)&\to
\begin{cases}
 0 & r\ll l,
\\
\text{finite} & r\sim l,
\\
 0 & r\gg l,
\end{cases}
\end{align}
from which the characteristic scale \(l\) is identified.
In practical situations, the scale susceptibility \(G_{q}(r)\) is more
convenient than \(F_{q}(r)\) to identify the characteristic scale \(l\).
We demonstrate how \(F_{q}(r)\) and \(G_{q}(r)\) work by applying them
to synthetic images with a definite characteristic scale and different
levels of concentration fluctuation.
We consider
the model systems in Fig.~\ref{fig:dichotomous_tonedown_pattern}.
Each system in Fig.~\ref{fig:dichotomous_tonedown_pattern} consists of
168\(\times\)168 pixels, which define the range of \(r\) in the unit of
pixel is from 1 to 168.
and the concentration of a component is represented in gray-scale where
the value of concentration, \(c(\vec{x})\), is within [0,255].
The single characteristic scale \(l\) of each system is naturally defined as
the linear sizes of the darker areas, which are
84, 28, 7 and 2 pixels from the left to the right columns.
The systems in the first row model the strongly segregated mixtures
where the dichotomous values of concentration are 255 and 0.
By moving down the matrix, the concentration fluctuation becomes weaker
as the contrast between ``black'' and ``white'' reduces.

\begin{figure}[htbp]
\begin{minipage}[cbt]{\hsize}
 \begin{minipage}[cbt]{.49\hsize}
  \includegraphics[width=1.4\hsize]{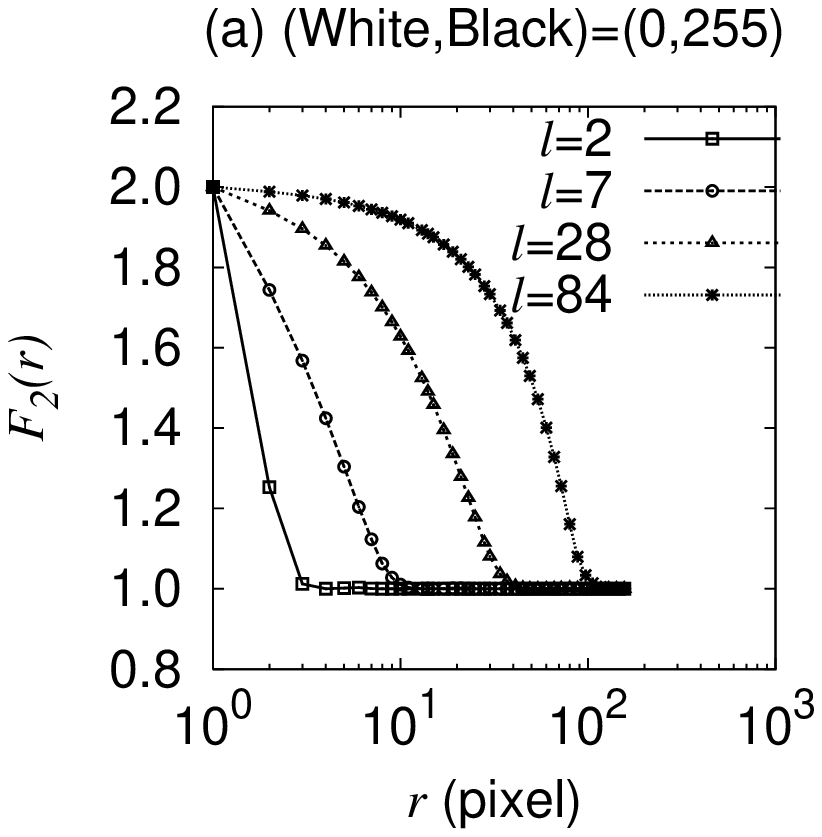} 
 \end{minipage}
\begin{minipage}[cbt]{.49\hsize}
  \includegraphics[width=1.4\hsize]{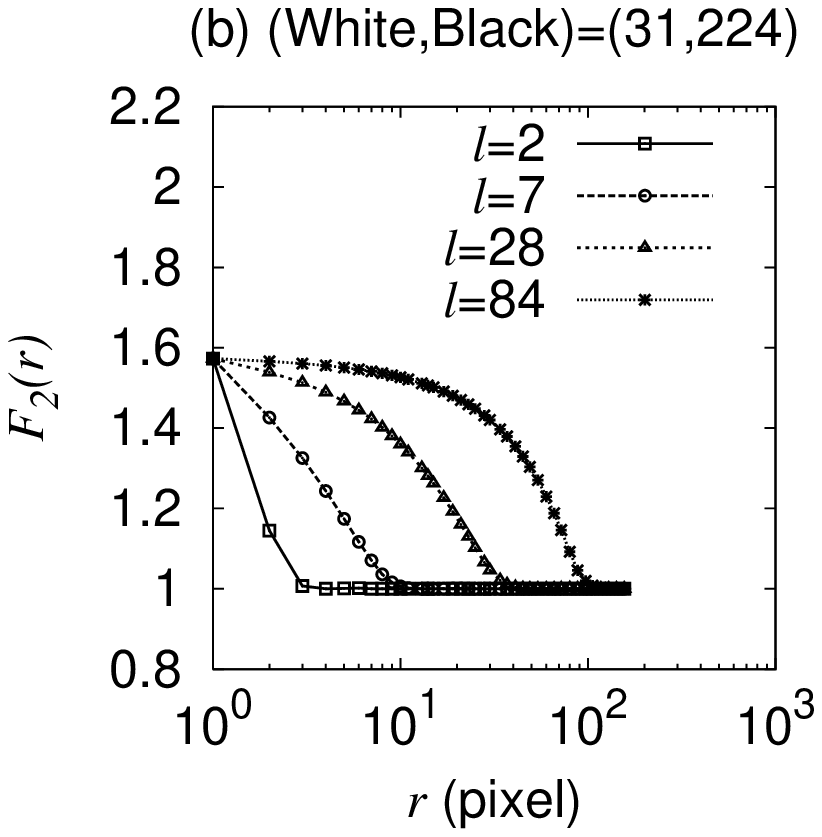} 
\end{minipage}
\\
\begin{minipage}[cbt]{.49\hsize}
  \includegraphics[width=1.4\hsize]{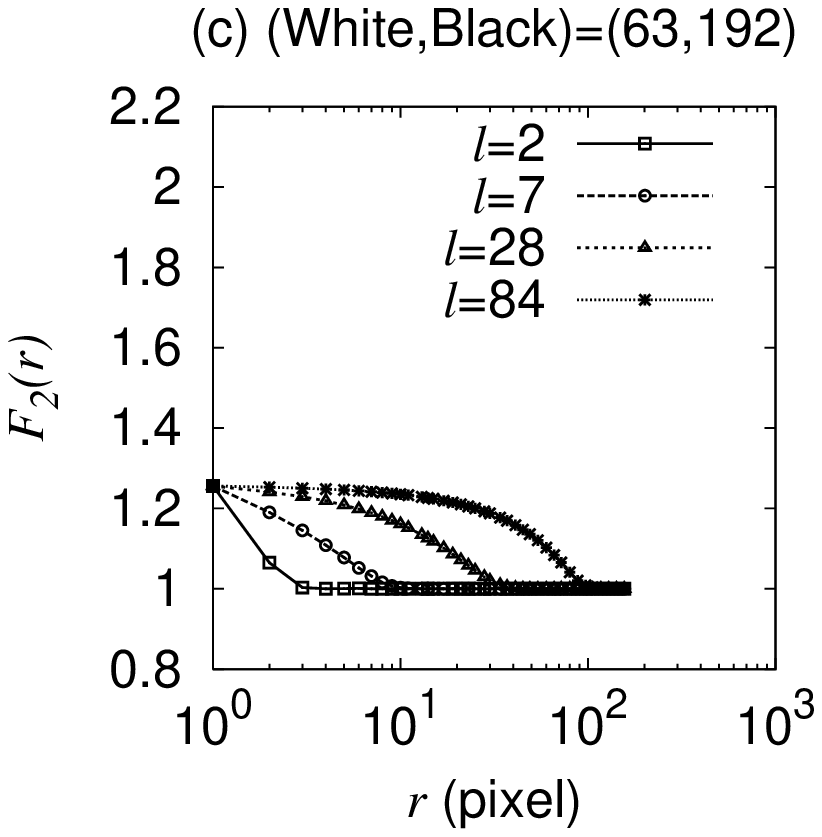} 
\end{minipage}
\begin{minipage}[cbt]{.49\hsize}
  \includegraphics[width=1.4\hsize]{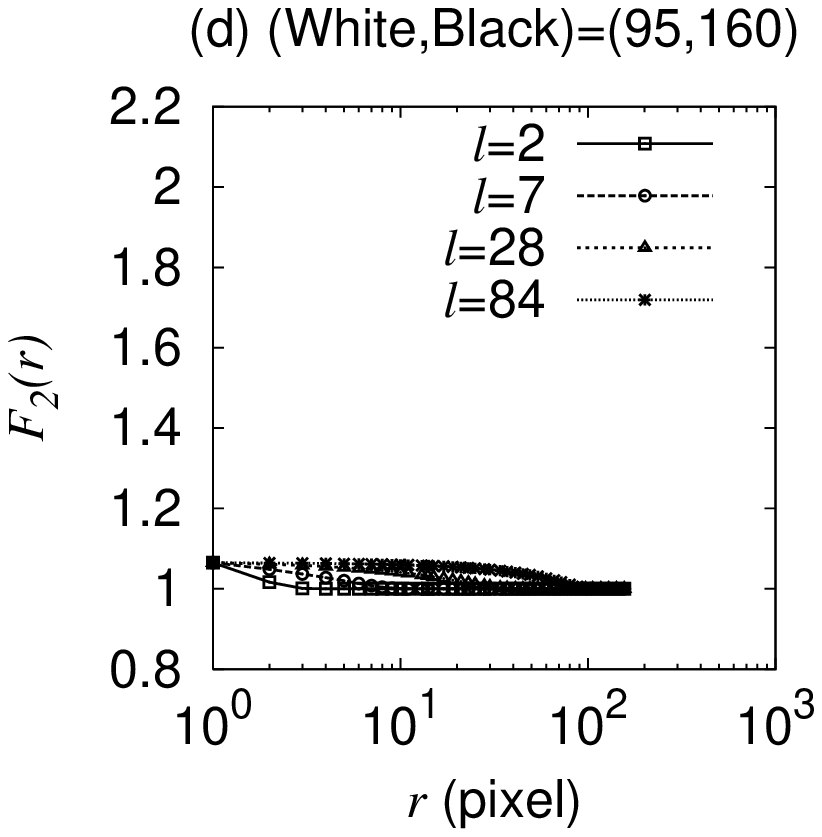} 
\end{minipage}
\\
\begin{minipage}[cbt]{.49\hsize}
  \includegraphics[width=1.4\hsize]{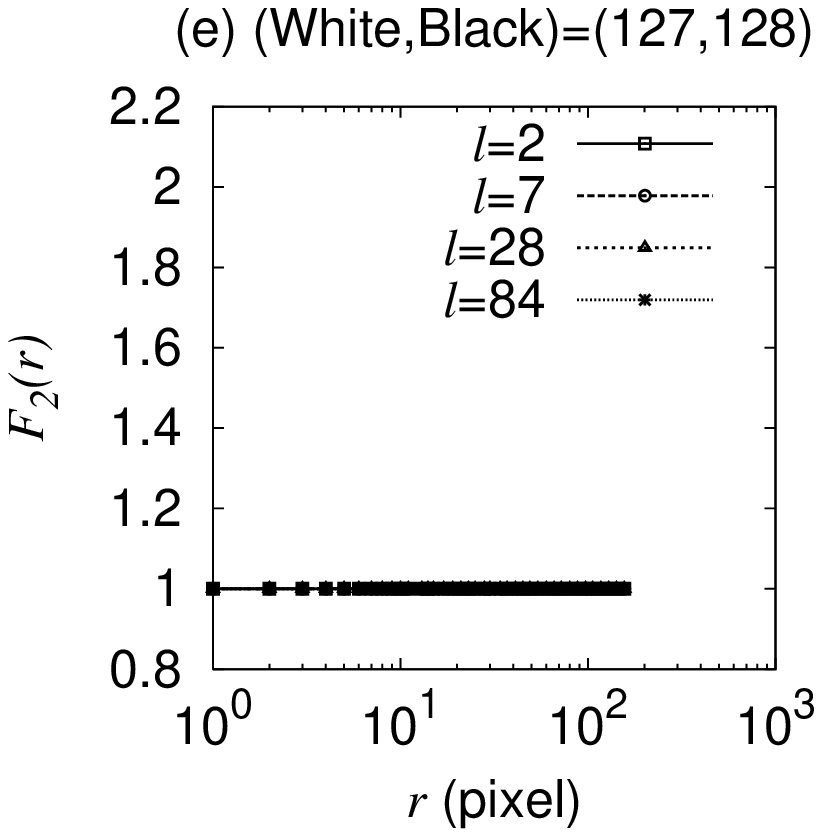} 
\end{minipage}
\begin{minipage}[cbt]{.49\hsize}
~ 
\end{minipage}
\end{minipage}
\caption{Inhomogeneity functions for the patterns in
Fig.~\ref{fig:dichotomous_tonedown_pattern} as a function of the
resolution scale. The ``black'' and ``white'' intensities are (a) 0:255,
(b) 31:224, (c) 63:192, (d) 95:160 and (e) 127:128.}
\label{fig:moment_dichotomous_tonedown_pattern}
\end{figure}

The inhomogeneity functions of \(q=2\) for the systems in
Fig.~\ref{fig:dichotomous_tonedown_pattern} are drawn in
Fig.~\ref{fig:moment_dichotomous_tonedown_pattern}.
Because the concentration value is dichotomous in these systems,
a single value of \(q\) is sufficient. Different values of \(q\) did not change the
qualitative aspect of \(F_{q}(r)\).
The \(F_{q}(r)\) for all the systems shows the sharp variation
in each characteristic scale.
The differences in the concentration fluctuation are explained by the
magnitude of \(F_{q}(r)\) at \(r\ll l\); the stronger the
concentration fluctuation, the larger the inhomogeneity function.
For this dichotomous-valued field, an \(F_{q}(r\to 1)\) is at the finest
resolution and is expressed as
\(
 F_{q}(1)
= 2\left[
\left\{
1-
\left(
\frac{w}{255}
\right)
\right\}^{q}+
\left(
\frac{w}{255}
\right)
^{q}
\right]^{1/(q-1)},
\) where \(w\) denotes the intensity of ``white'' region.
This result shows that the intensity of segregation is characterized by
the value of \(F_{q}(r)>1\).

\begin{figure}[htbp]
\begin{minipage}[cbt]{\hsize}
 \begin{minipage}[cbt]{.42\hsize}
\includegraphics[width=1.0\hsize]{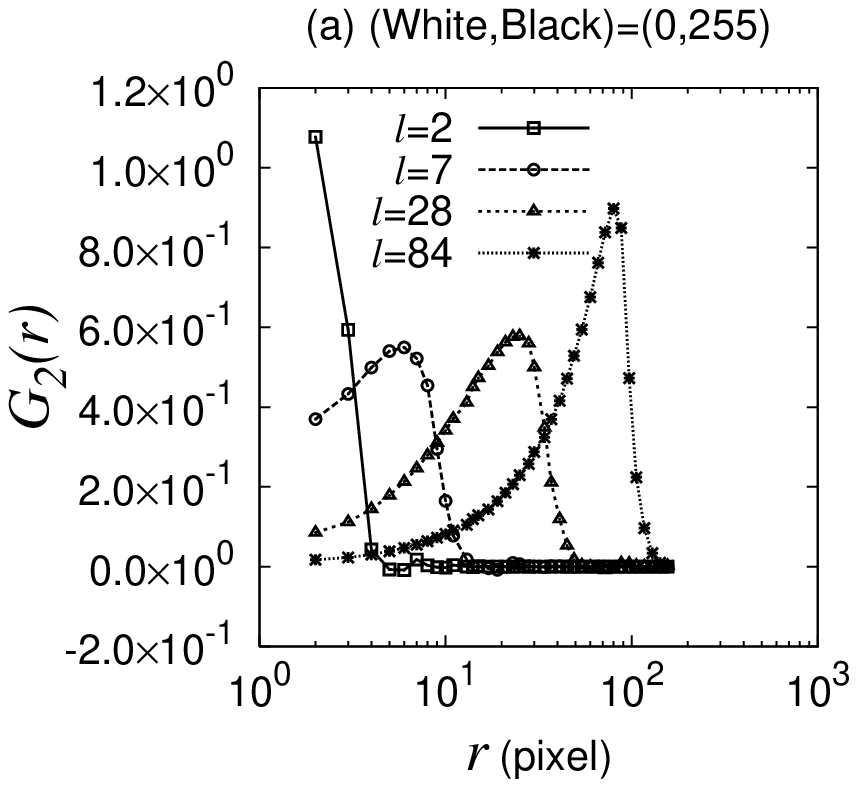}
 \end{minipage}
\begin{minipage}[cbt]{.42\hsize}
\includegraphics[width=1.0\hsize]{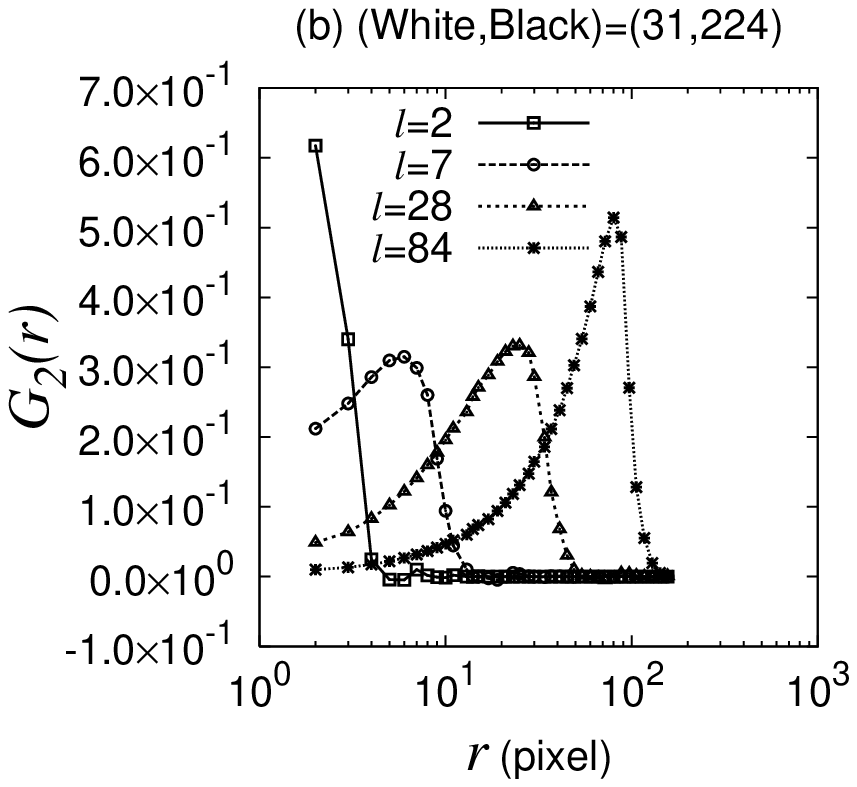}
\end{minipage}
\\
\begin{minipage}[cbt]{.42\hsize}
\includegraphics[width=1.0\hsize]{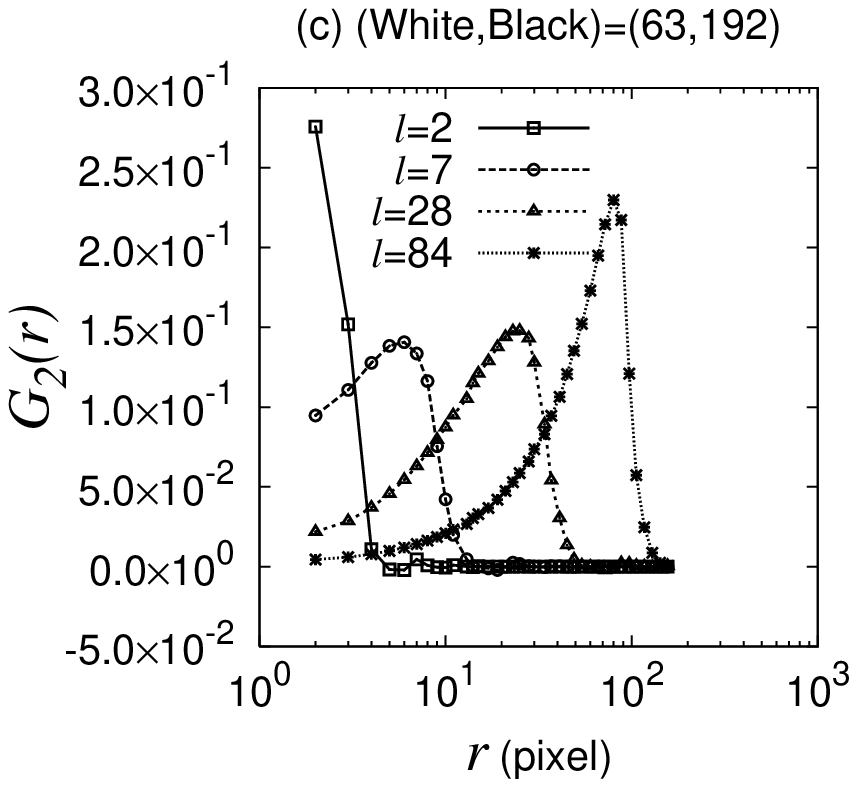}
\end{minipage}
\begin{minipage}[cbt]{.42\hsize}
\includegraphics[width=1.0\hsize]{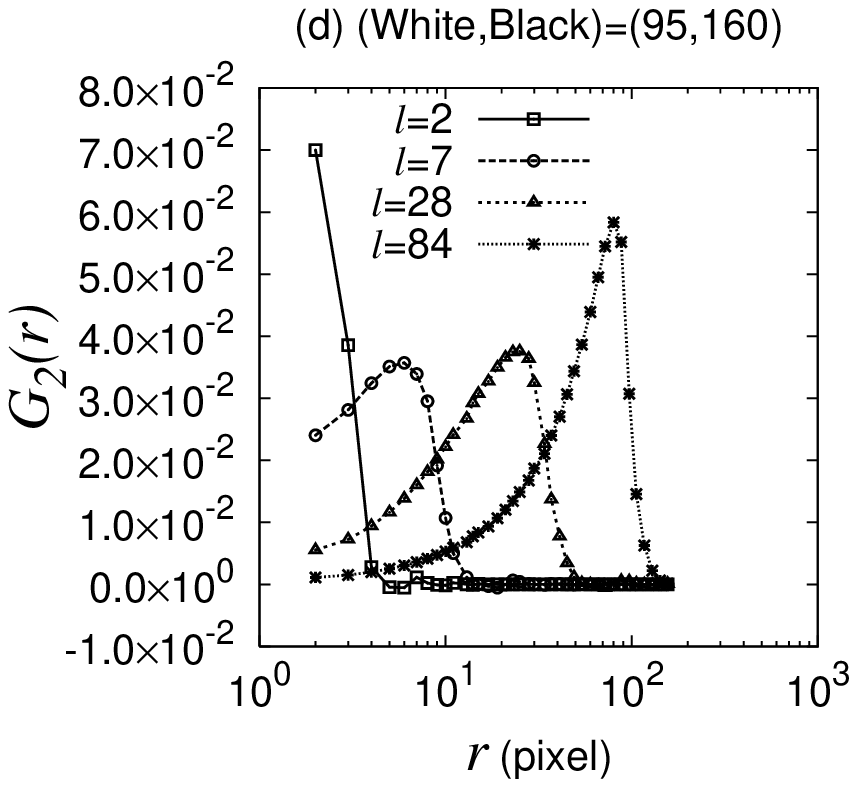}
\end{minipage}
\\
\begin{minipage}[cbt]{.42\hsize}
\includegraphics[width=1.0\hsize]{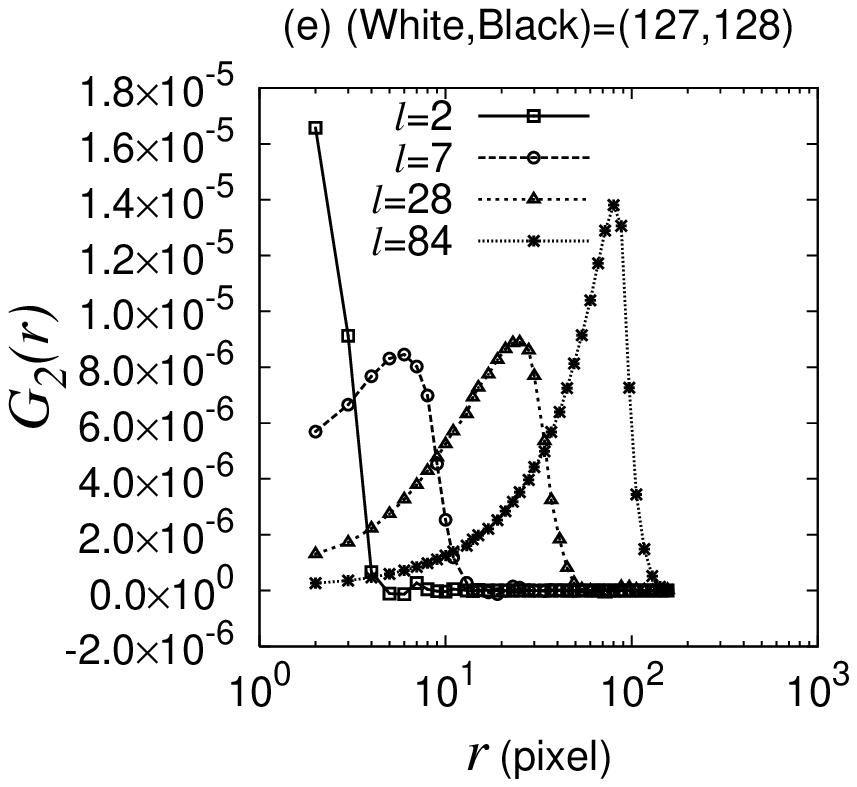}
\end{minipage}
\begin{minipage}[cbt]{.42\hsize}
~ 
\end{minipage}
\end{minipage}
\caption{Scale susceptibility to the resolution scale, \(G_{2}(r)\), for
the systems in Fig.~\ref{fig:dichotomous_tonedown_pattern} as a function
of the resolution scale.  The ``black'' and ``white'' intensities are
(a) 0:255, (b) 31:224, (c) 63:192, (d) 95:160 and (e) 127:128.}
\label{fig:dmoment_dichotomous_tonedown_pattern}
\end{figure}

The characteristic scale and intensity of segregation are independent
characteristics of a mixture and are simultaneously accounted for by
\(F_{q}(r)\).
To isolate the characteristic scales, the scale susceptibility
\(G_{2}(r)\) for patterns in Fig.~\ref{fig:dichotomous_tonedown_pattern}
is shown in Fig.~\ref{fig:dmoment_dichotomous_tonedown_pattern}.
The \(G_{2}(r)\) for the same scale of segregation is almost identical,
except for the ordinate scale, and clearly indicates \(l\).
The usefulness of \(G_{2}(r)\) is for systems with small
fluctuations but a definite geometric structure, such as the systems of
(``white'',``black'')=(127:128) in
Fig.~\ref{fig:dichotomous_tonedown_pattern}.
The inhomogeneity functions in
Fig.~\ref{fig:moment_dichotomous_tonedown_pattern}(e) for these
(127:128)-systems are close to unity at all scales, and the
characteristic scales are almost indiscernible without magnification.
However, the peaks in \(G_{2}(r)\) in
Fig.~\ref{fig:dmoment_dichotomous_tonedown_pattern}(e)
clearly show the \(l\) even for this weakly segregated system.

We add some comments on the relationship between 
mixture quantification without accounting for a scale
and the method of the scale-dependent moment.
In the limit of \(r\to 0\), the moment function of a coarse-grained
concentration becomes \(\langle \mu_{r\to 0}^{q}\rangle \to
\int_{\Omega}\upd\vec{x}\mu^{q}(\vec{x})\), which is proportional to the
usual moment of the concentration field.
The variance-based measures, such as the intensity of segregation, are
basically related to this limit of \(q=2\).
Such quantifications describe some average deviation from the uniform
distribution in the smallest resolution and therefore are insensitive to
the scale of any internal structure.

\section{Results and discussion}
\subsection{Application to particle dispersions}
A particle dispersion and a composite material are typical examples of
a multi-component system, which consists of small particles and a
matrix medium.
The physical properties of a dispersion are not solely determined by
those of the particles and the matrix medium but also depend on the
distribution of the particles.
The structure of the particle distribution is controlled by the
non-equilibrium processing history, as well as by the thermodynamic
stability of the structure.
The mixture state of a dispersion should not be characterized
solely by the primary particle size but also by multiple scales associated
with particle distributions.
We apply the moment function method for characterization of different
states of synthetic particle dispersions.

\begin{figure}[htbp]
 \center
\begin{minipage}{\hsize}
\begin{minipage}[cbt]{.32\hsize}
(a)
\\
 \includegraphics[width=\hsize]{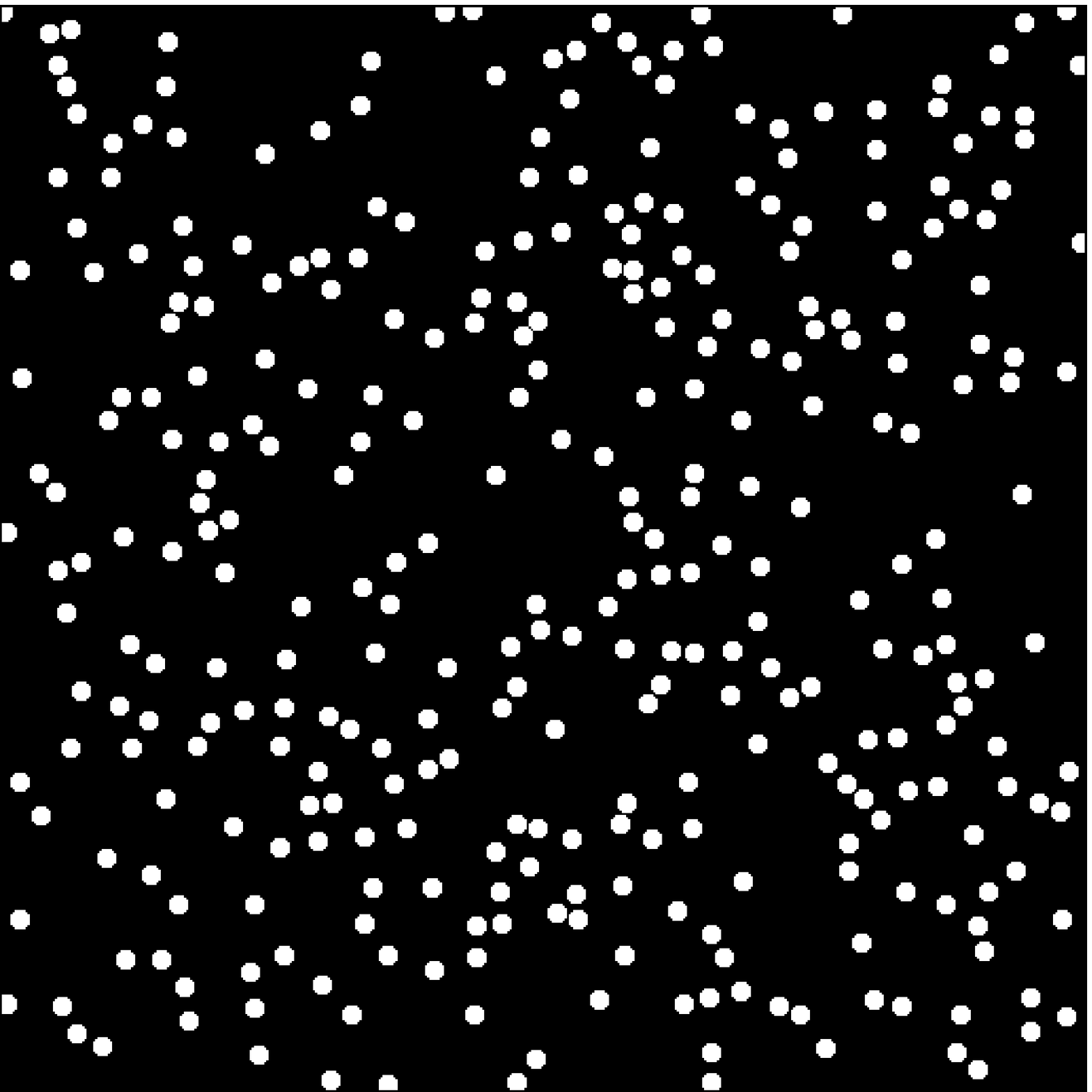}
\end{minipage}
\begin{minipage}[cbt]{.32\hsize}
(b)
\\
 \includegraphics[width=\hsize]{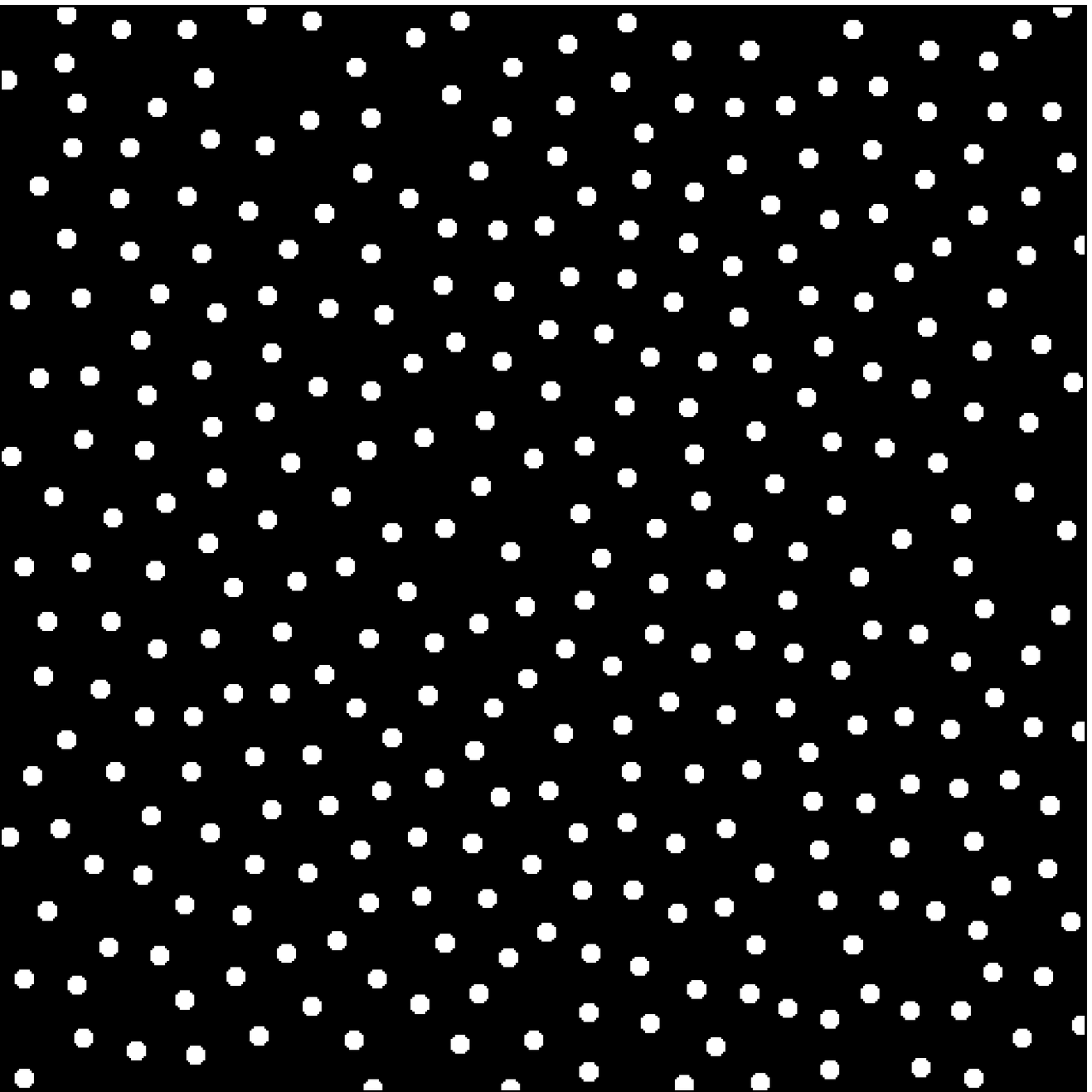}
\end{minipage}
\begin{minipage}[cbt]{.32\hsize}
(c)
\\
 \includegraphics[width=\hsize]{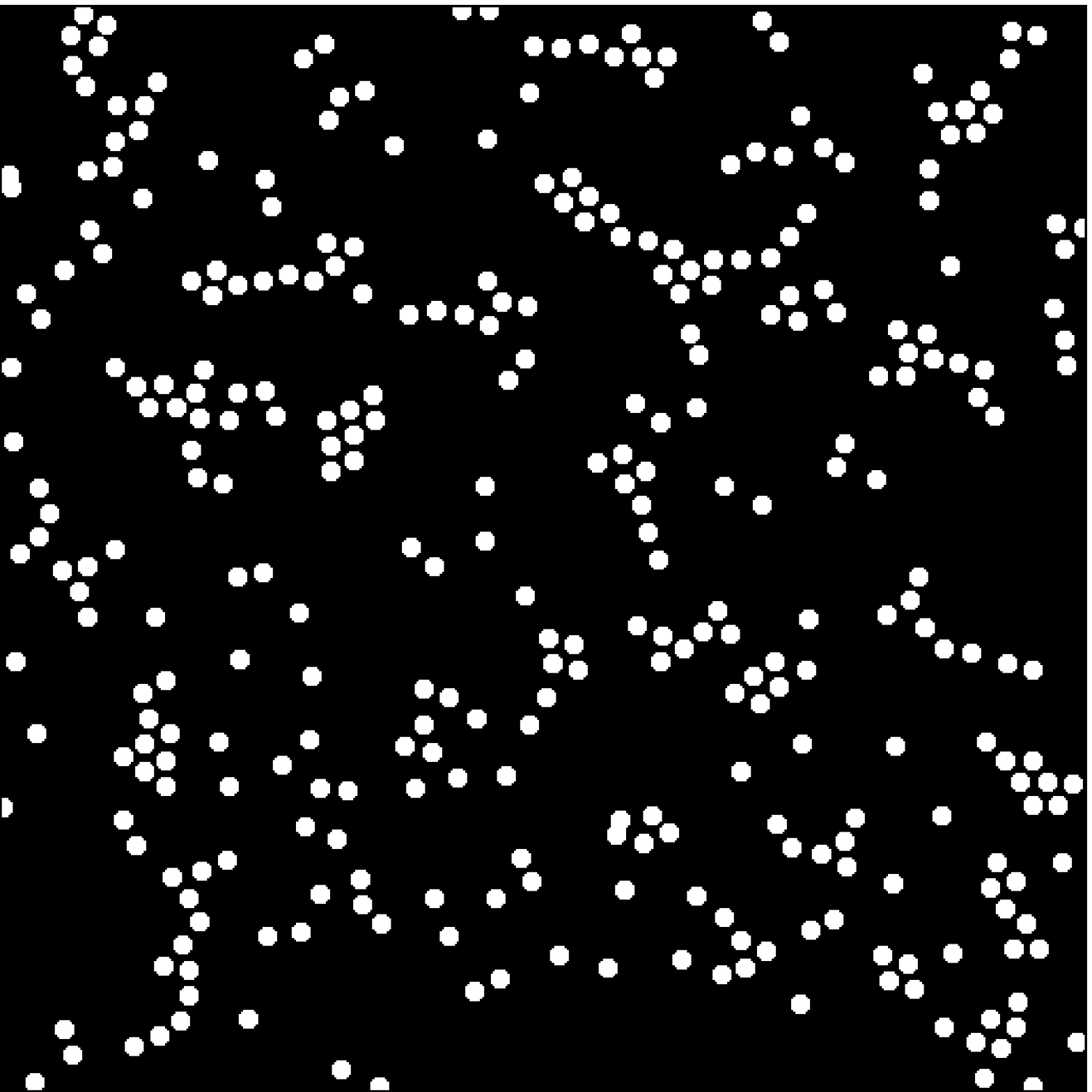}
\end{minipage}
\end{minipage}
\\
\vspace*{2ex}
\begin{minipage}[cbt]{\hsize}
\begin{minipage}[cbt]{.49\hsize}
(d)
\\
 \includegraphics[width=\hsize]{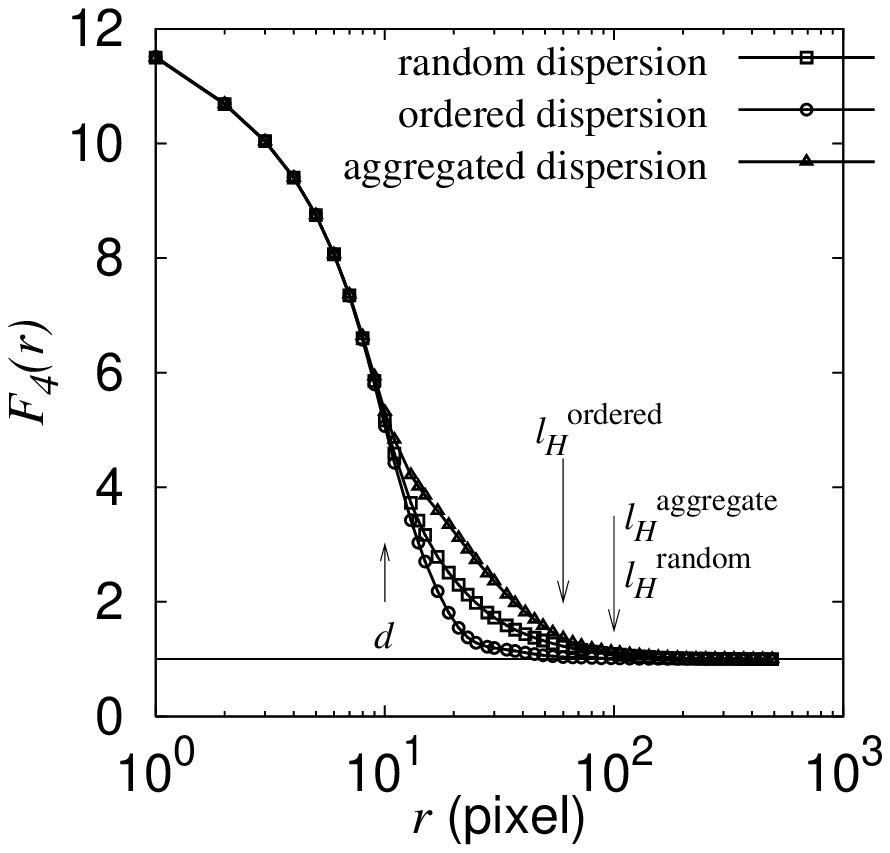}
\end{minipage}
\begin{minipage}[cbt]{.49\hsize}
(e)
\\
 \includegraphics[width=\hsize]{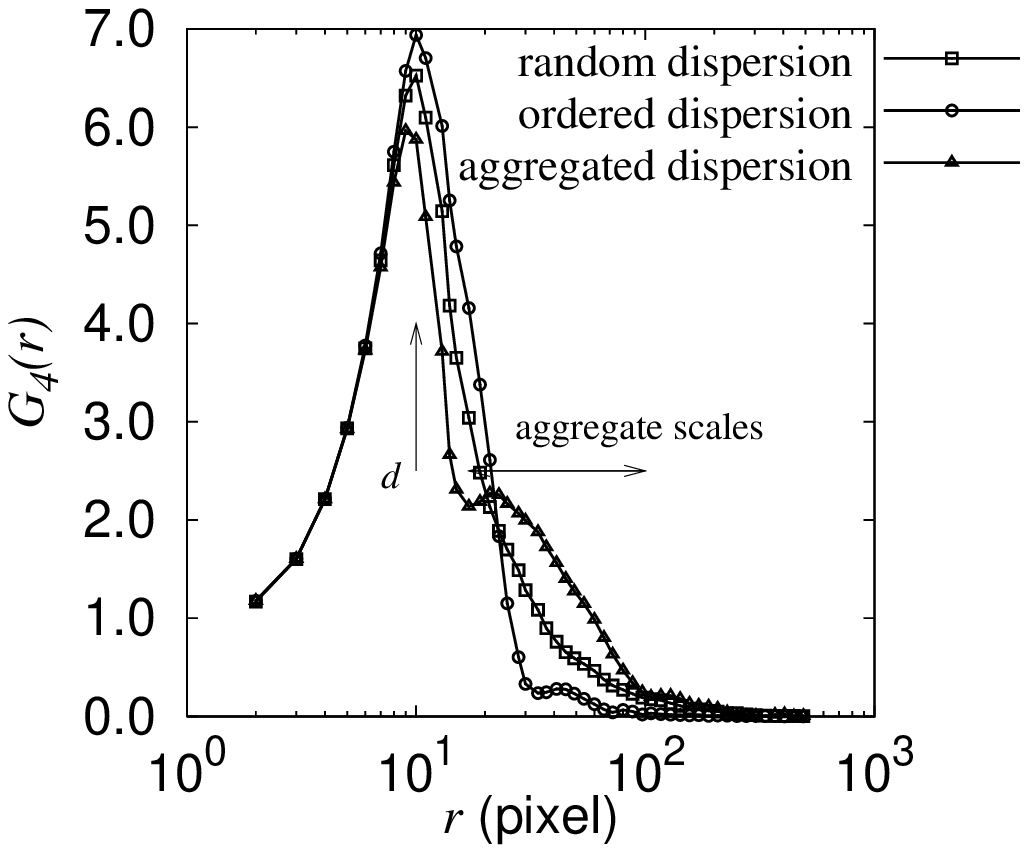}
\end{minipage}
\end{minipage}

\begin{minipage}[cbt]{\hsize}
\begin{minipage}[cbt]{.32\hsize}
(f)
\\
 \includegraphics[width=\hsize]{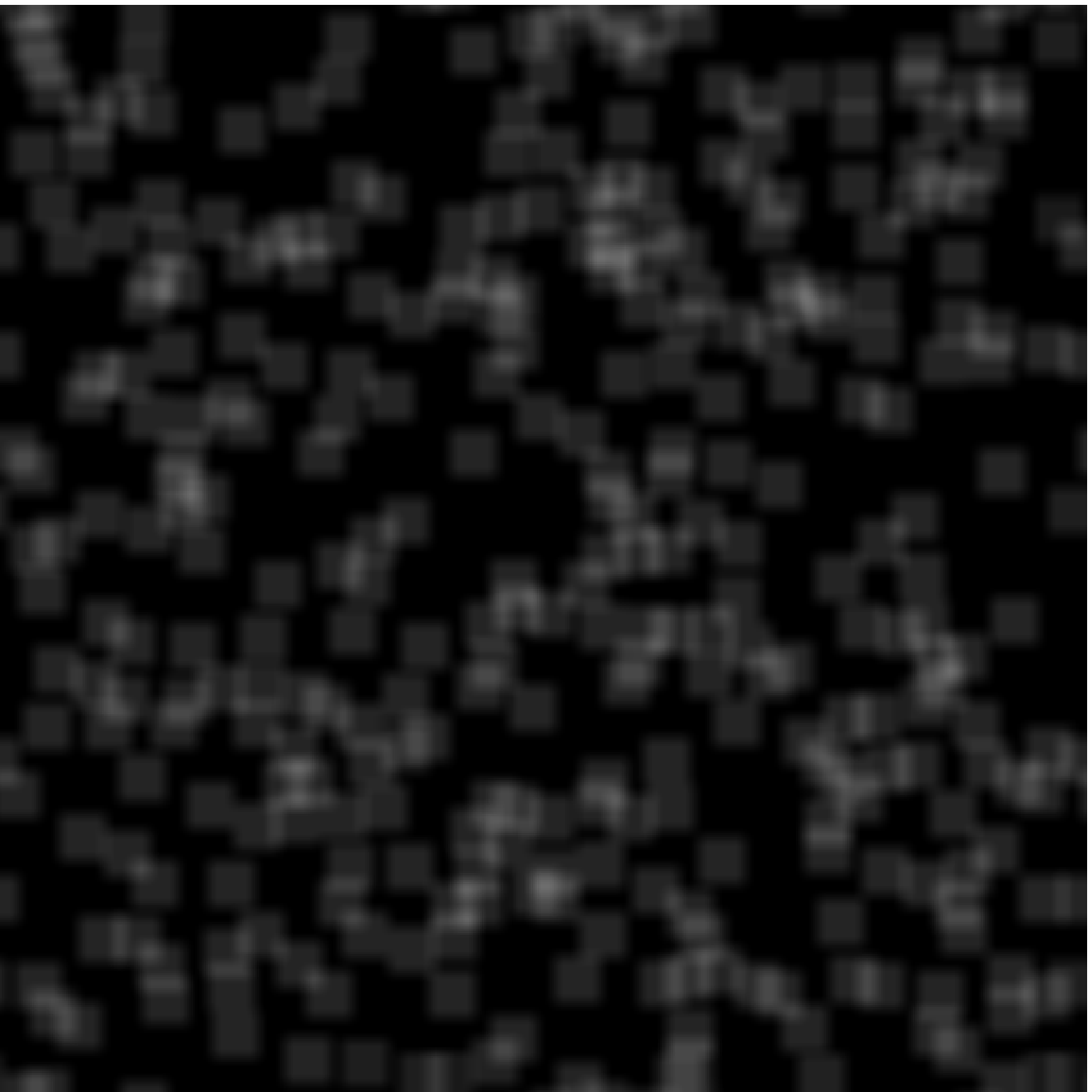}
\end{minipage}
\begin{minipage}[cbt]{.32\hsize}
(g)
\\
 \includegraphics[width=\hsize]{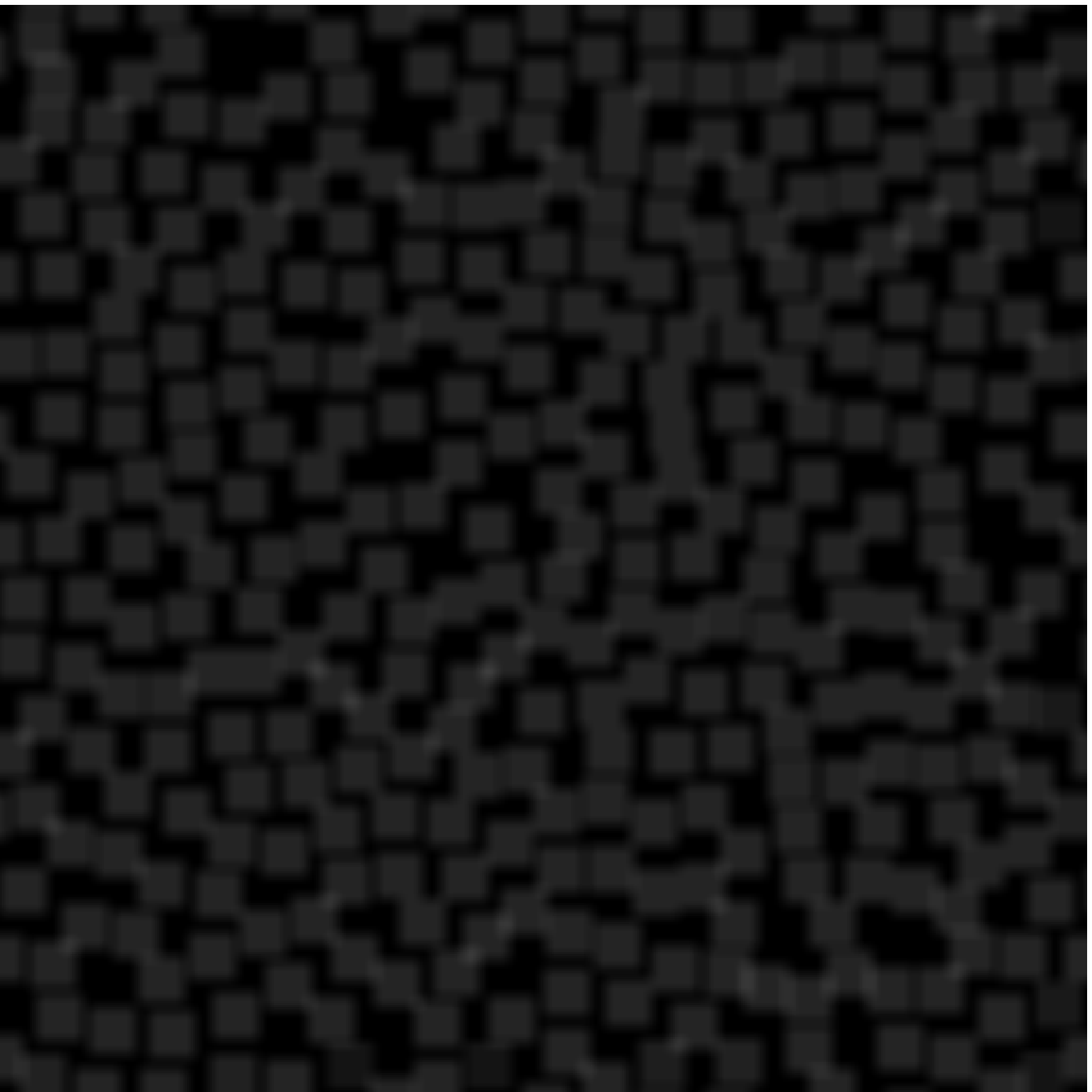}
\end{minipage}
\begin{minipage}[cbt]{.32\hsize}
(h)
\\
 \includegraphics[width=\hsize]{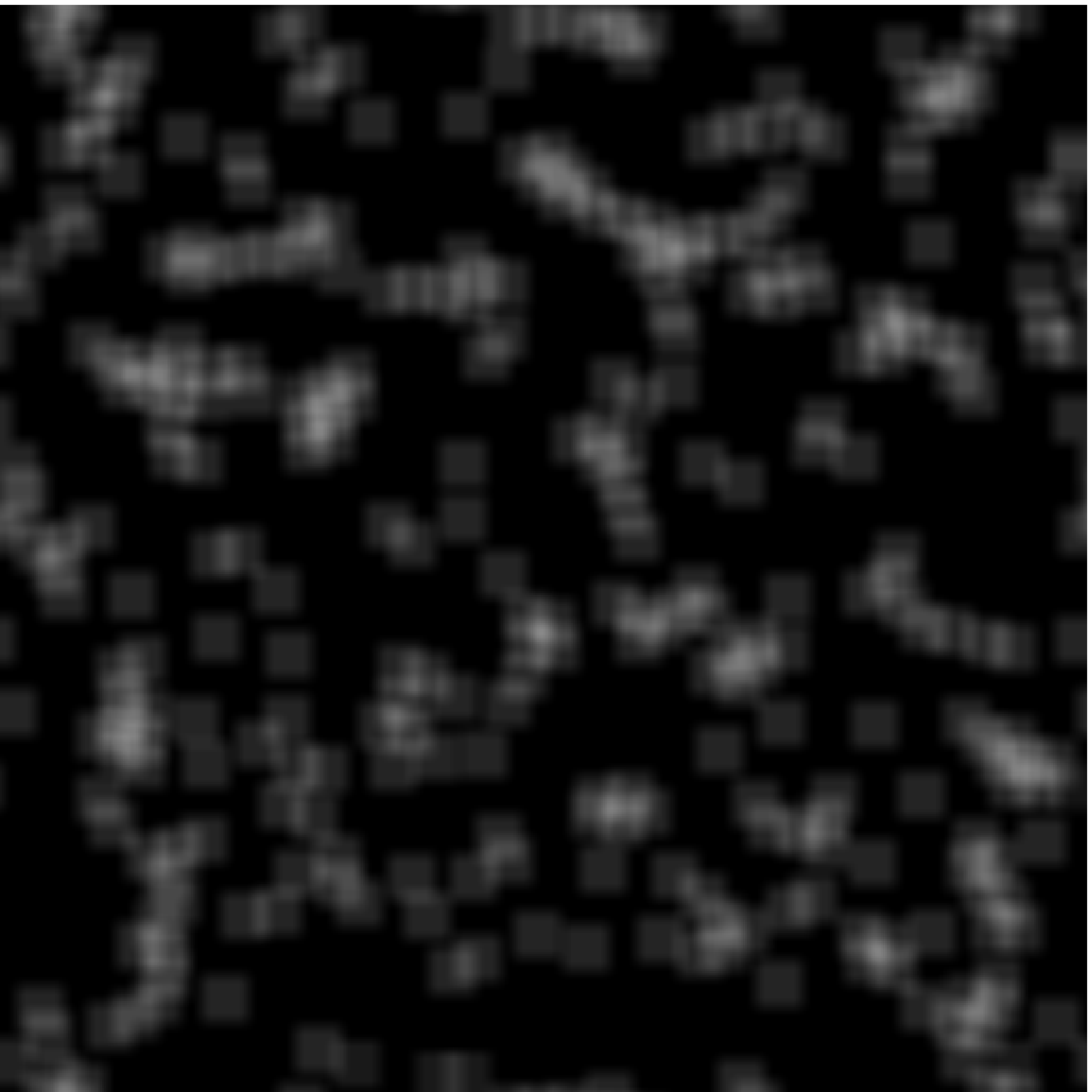}
\end{minipage}
\end{minipage}
\caption{Particle dispersion and their multiple-scale inhomogeneity. 
Three different particle dispersions of 512\(\times\)512
pixels: (a) particles are randomly distributed, (b) particles are
locally ordered, and (c) particles are aggregated.
The area fraction of the systems is 0.1 and 
the diameter of the particles is \(d=10\)  pixels. 
(d)~The inhomogeneity functions, \(F_{4}(r)\), and the scale
susceptibilities, \(G_{4}(r)\), for the systems~(a)-(c).
Coarse-grained fields, \(\mu_{22}(\vec{x})\), for
(f) random dispersion, (g) ordered dispersion and (h) aggregated dispersion.
}
\label{fig:dispersion_pattern}
\vspace*{1ex}
\end{figure}

Consider monodisperse dispersions with a diameter of the particles
\(d=10\)~pixels that is a natural characteristic scale of the systems.
For the sake of simplicity, two-dimensional systems are analyzed, but
application to three-dimensional systems is straightforward.
Three different dispersions with an area fraction of the particles of 0.1 are
depicted in Fig.~\ref{fig:dispersion_pattern}(a)-(c). The particles are randomly
distributed in Fig.~\ref{fig:dispersion_pattern}(a), the particles are
locally ordered in Fig.~\ref{fig:dispersion_pattern}(b) and several
particles aggregate in Fig.~\ref{fig:dispersion_pattern}(c).
The inhomogeneity function, \(F_{4}(r)\), and the scale susceptibility,
\(G_{4}(r)\), for the three systems in
Figs.~\ref{fig:dispersion_pattern}(a)-(c) are shown in
Fig.~\ref{fig:dispersion_pattern}(d)
and Fig.~\ref{fig:dispersion_pattern}(e), respectively.
In this case, the concentration field of the particles was analyzed,
 which was unity in the particle domain and zero in the matrix domain.
Because the analysis was applied to binarized systems, a single value of
\(q\) is sufficient. For clarity of presentation, the results of \(q=4\) are shown.
For the random dispersion in Fig.~\ref{fig:dispersion_pattern}(a), 
there are no coherence scales other than \(d\) because the particles are
randomly distributed by construction.
The \(F_{4}(r)\) and 
\(G_{4}(r)\) for the random dispersion indicate the particle size by the
large fluctuation at \(r=d\) and the scale of homogeneity for
\(l_{H}^{\text{random}}\approx 10d\).
The ratio \(l_{H}^{\text{random}}/d=10\) is relatively large, which
explains the inhomogeneity of the particle distribution.

In the system of Fig.~\ref{fig:dispersion_pattern}(b), inter-particle
distances between nearest pairs are almost the same, and the particles
are locally ordered; therefore we call the system an ordered dispersion for
convenience.
The average inter-particle distance is \(r_{1}\approx 2.7d\), which was
measured by the first peak in the correlation function or the pair
correlation function. 
The \(r_{1}\) and \(d\) are the natural characteristic scales 
in the ordered dispersion.
The large fluctuation at \(r=d\) in \(F_{4}(r)\) and 
\(G_{4}(r)\)
for the ordered dispersion indicates the particle size.
At \(r>r_{1}\) in \(F_{4}(r)\), the fluctuation almost vanishes because
the structural fluctuation does not exists at large scale.
A close examination of \(G_{4}(r)\) reveals a small peak at around
\(r\approx 4d\), which can be ascribed to the fluctuation in a
middle-range scale.
The scale of homogeneity is about \(l_{H}^{\text{ordered}}\approx 6d\).
If the ordered configuration developed globally, 
\(l_{H}^{\text{ordered}}\approx r_{1}\) would be expected because the
fluctuation \(r>r_{1}\) did not exist.
The ordered configuration makes the scale of homogeneity smaller than that
of the random dispersion.

In the system of Fig.~\ref{fig:dispersion_pattern}(c), several particles
aggregate, and these aggregates are distributed randomly
in an aggregated dispersion.
One aggregate is composed of 2-20 particles, thus the aggregated
dispersion should be characterized by the multiple sizes of the
aggregates and the nearest-neighbor distance \(r_{1}=1.2d\) in an
aggregate besides the particle size \(d\).
In addition to the large fluctuation at \(r=d\) by the particle size,
another broad band at \(r_{1}<r<10d\) in \(G_{4}(r)\) are observed,
which corresponds to the multiple sizes of the aggregates.
The scale of homogeneity is about
\(l_{H}^{\text{aggregate}}\approx 10d\), which is similar to that in the
random dispersion and much greater than that in the ordered dispersion
because of the random distribution of the aggregates.
Difference of mixture quality among three dispersions
is obvious in the scale susceptibility function, \(G_{4}(r)\), in
Fig.~\ref{fig:dispersion_pattern}(e).
Identification of multiple characteristic scales based on inhomogeneity
is essential to quantify mixture quality.

Figures~\ref{fig:dispersion_pattern}(f)-(h) show the coarse-grained
fields of the particle dispersions in
Figs.~\ref{fig:dispersion_pattern}(a)-(c) with the resolution scale of
\(r=2.2d\).
At this resolution scale, a single particle is hardly discernible.
In Fig.~\ref{fig:dispersion_pattern}(b), the \(\mu_{22}(\vec{x})\) for
the ordered dispersion, which has the smallest \(l_{H}\), shows the
smallest contrast indicating the smallest fluctuation.
For the random dispersion, \(\mu_{22}(\vec{x})\) shows a random
structure in the fluctuation~(Fig.~\ref{fig:dispersion_pattern}(a)).
For the aggregated dispersion, 
\(\mu_{22}(\vec{x})\) shows a streak structure and the largest
fluctuation among the three systems~(Fig.~\ref{fig:dispersion_pattern}(c)).

Coarse-graining is a key concept 
in simultaneous characterization of 
various
characteristic scales associated with a hierarchical structure and
inhomogeneity of distribution.
The demonstration with the different dispersions above showed how the
scale-dependent moment method works to assess different characteristic
scales including the particle size, aggregate sizes, and inter-particle
spacing.

\section{Concluding remarks}
We proposed a method for characterization of hierarchical mixture quality
of complex mixtures , which we call the scale-dependent moment function
method.
In this method, the inhomogeneity of a mixture is defined at different
scales.
From this scale-dependent inhomogeneity, multiple characteristic scales in
mixture quality are identified.
In other words, 
the sizes and distributions of internal structures are simultaneously
accounted for in the scale-dependent moment function.
Application of this method to dispersions with different structures showed that
differences in mixture quality are characterized by multiple length scales
and associated inhomogeneity.

The scale-dependent moment method is applicable to two- or
three- dimensional concentration fields.
The mixture quality characterization based on the scales and
inhomogeneity should be effective not only for multi-component complex
mixtures with hierarchical internal structures but also for the evolution of
the mixing process of both miscible and immiscible mixtures.
It is expected that this proposed multiple-scale characterization can provide an
insight on the relationship between the distribution of internal structures and
the physical properties of mixture materials, but this issue is left for
future work.

%

%

%


\end{document}